# Photo Thermal Effect Graphene Detector Featuring 105 Gbit s$^{-1}$ NRZ and 120 Gbit s$^{-1}$ PAM4 Direct Detection

S. Marconi[1], M. A. Giambra[2], A. Montanaro[2], V. Mišeikis[3,4], S. Soresi[2,5], S. Tirelli[2,5], P. Galli[6], F. Buchali[7], W. Templ[7], C. Coletti[3,4], V. Sorianello[2] and M. Romagnoli[2,*]

[1]Tecip Institute – Scuola Superiore Sant'Anna, Via G. Moruzzi 1, 56124 Pisa, Italy

[2]Photonic Networks and Technologies Lab – CNIT, Via G. Moruzzi 1, 56124 Pisa, Italy

[3]Center for Nanotechnology Innovation @NEST - Istituto Italiano di Tecnologia, Piazza San Silvestro 12, I-56127 Pisa, Italy

[4]Graphene Labs, Istituto Italiano di Tecnologia, Via Morego 30, 16163 Genova, Italy

[5]Fondazione INPHOTEC, Via G. Moruzzi 1, 56124 Pisa, Italy

[6]Nokia Solutions and Networks Italia, via Energy Park 14, 20871 Vimercate, Italy

[7]Nokia Bell Labs, Lorenzstr. 10, 70435 Stuttgart, Germany

*corresponding author: marco.romagnoli@cnit.it

The challenge of next generation datacom and telecom communication is to increase the available bandwidth while reducing the size, cost and power consumption of photonic integrated circuits[1,2]. Silicon (Si) photonics has emerged as a viable solution to reach these objectives[3,4]. Graphene, a single-atom thick layer of carbon[5], has been recently proposed to be integrated with Si photonics because of its very high mobility[6], fast carrier dynamics[7] and ultra-broadband optical properties[8]. Here, we focus on graphene photodetectors for high speed datacom and telecom applications. High speed graphene photodetectors have been demonstrated so far[9], however the most are based on the photo-bolometric (PB)[10–12] or photo-conductive (PC)[13,14] effect. These devices are characterized by large dark current, in the order of milli-Amperes, which is an impairment in photo-receivers design[15], Photo-thermo-electric (PTE) effect has been identified as an alternative phenomenon for light detection[16]. The main advantages of PTE-based photodetectors are the optical power to voltage conversion, zero-bias operation and ultra-fast response[17]. Graphene PTE-based photodetectors have been reported in literature[18–22], however high-speed optical signal detection has not been shown. Here, we report on an optimized graphene PTE-based photodetector with flat frequency response up to 65 GHz. Thanks to the optimized design we demonstrate a system test leading to direct detection of 105 Gbit s$^{-1}$ non-return to zero (NRZ) and 120 Gbit s$^{-1}$ 4-level pulse amplitude modulation (PAM) optical signals.

The global number of devices connected to the IP network is growing at a compound annual growth rate (CAGR) of 10%, and is expected to exceed 28 billion connected devices by 2023[23,24]. The speed of these connections will also increase, with 5G devices expected to reach 575 Mbps by 2023[23]. As a result, next optical network and interconnect technologies will be developed to meet the increasing bandwidth connectivity demands. One of the next major transitions will be the general availability of 400-Gigabit Ethernet (GbE) technology[24], with optical transceivers providing higher capacity at a reduced cost, footprint, and power consumption.

In this scenario, Si photonics technology plays an essential role as it offers the possibility of cost-effective large volume production, thanks to the availability of well-established silicon fabrication facilities and the relatively low cost and high abundance of the material[3,4]. Si photonics transceivers have reached the market and with a potential growth at a CAGR exceeding 20% according to market analysts[25,26]. Typical Si photonics transceivers may include: modulators based on Si junctions realized by implantation doping[27], modulators based on germanium (Ge) or silicon-germanium (SiGe) epitaxy[28,29], photodetectors based on Ge epitaxy[30].

2D materials have recently emerged as viable alternative enablers for Si photonics devices[31]. In particular, graphene has been recognized as a promising material to be integrated on Si photonics to fulfil the requirements of next generation transceivers for datacom and telecom applications[17]. Graphene is an allotrope of carbon with atoms arranged on a one-atom thick hexagonal lattice[5]. The peculiar atomic structure, made of covalent bonds, determines a linear dispersion between energy and momentum with the conduction and valence bands meeting in single point (Dirac point) in the momentum space, with no energy gap[7]. This feature leads to intriguing electronic and optical properties, for example, ultra-high carrier mobility[6] and tuneable electronic and optical properties by electric field gating[32,33]. Moreover, graphene can be grown on a sacrificial material by CVD[34] and then transferred on top of any fully passive photonic waveguide platform, e.g. Si or Silicon Nitride (SiN), in order to realize broadband photodetectors[9], fast electro-absorption[35] and electro-refraction[36] modulators with a potential impact on the cost of future graphene photonic transceivers[17].

Graphene photodetectors offer several advantages over the typical Ge photodetectors[30]. The first is the ultra-broadband optical absorption of graphene from the UV to the far infrared and beyond, which is due to the band structure with zero bandgap at the Dirac point[37]. Moreover, the transport properties of graphene characterized by ultra-high mobility and fast dynamics of hot carriers (HCs) upon optical excitation, with relaxation times of the order of few ps[38–40], enables optical to electrical conversion with bandwidths exceeding hundreds of GHz[41,42].

Waveguide integrated graphene photodetectors with optoelectronic bandwidths ranging from some tens of GHz up to 128 GHz have been reported[10–14,18–22]. These devices are based on a single layer of graphene placed on top of an optical waveguide and provided with two metal contacts for photo-generated current or voltage extraction. Different photo-detection mechanisms are exploited to generate the signal: photo-bolometric (PB)[10–12], photo-conductive (PC)[13,14] and photo-thermo-electric (PTE) effect[18–22]. PB and PC based devices typically exhibit good current responsivity up to 0.5 AW$^{-1}$ [12], however they need a bias voltage to operate leading to a direct current (DC) that can be in the order of few mA[10–14]. In general, high DC currents may contribute to the overall noise because of the shot and generation recombination noise in graphene[43–45]. Moreover, photo-detectors are usually coupled to a trans-impedance amplifier (TIA) which converts the photocurrent into a voltage signal to a level that can be processed by the electronics downstream of the TIA[15]. In some applications, e.g. coherent detection, the DC current from the photodetectors is already in the order of few mA because of the DC photocurrent generated by the presence of the local oscillator[46]. In this scenario, the high dark DC current in PB and PC device may represent an issue causing the nonlinearity and saturation of the TIA[15]. In PTE based photodetectors, a photovoltage is generated through a thermoelectric effect induced by photon absorption[16]. PTE effect is highly efficient, as a large portion of the photon energy is transformed in electron heat owing to the ultra-fast carrier scattering and weak coupling to phonons in graphene[39,47]. In particular, absorbed photons generate HCs with a carrier temperature spatially distributed along the graphene layer following the optical intensity distribution[48]. By inducing a spatial profile of the Seebeck coefficient in the material, e.g. by electrical gating, an electromotive force is generated by thermoelectric effect (Seebeck effect) which is proportional to the spatial gradient of the HC's temperature[49–51]. The generated photovoltage does not require an externally applied bias, allowing zero dark current operation and direct voltage generation[48]. These properties are very appealing in terms of noise and power consumption, as PTE allows direct connection between the graphene photodetector and the read-out electronics without TIAs[17].

Waveguide integrated graphene detectors based on PTE effect have been reported in recent years showing voltage responsivity in the range 3.5 V W$^{-1}$-28 V W$^{-1}$ [18–22]. These devices are based on the enhancement of the gradient of HC's temperature through the improvement of the optical absorption by means of photonic structures confining the optical field in ~100nm gaps[18,19,22] or by exploiting the resonance of a microring resonator[20]. Despite the remarkable voltage responsivity[20,22], detection of an optical data transmission using PTE-based graphene photodetectors operating in unbiased condition has not been demonstrated yet. The optimization of the photovoltage signal in a PTE detector requires to set the detector at an operating condition near the Charge Neutrality Point (CNP), i.e. where the conductivity of graphene is lowest. For this reason, typical PTE photodetectors exhibit output resistance from several hundreds to thousands of Ohms[18–22], i.e. too high for proper matching to the typical 50 Ω impedance of the test instruments.

Here we report on a PTE graphene photodetector integrated on a Si photonic waveguide having a voltage responsivity of about 3.5 V W$^{-1}$ and a frequency response flat up to 65 GHz showing no roll off in the measured range. We optimized the design of the PTE detector to match 50 Ω resistance in order to be suitable for data communication system tests. We demonstrate for the first time the detection of high baud rate optical data stream by testing our device up to 105 GBaud Non-Return-to-Zero (NRZ) On-Off-Keying (OOK) and 60 GBaud 4-level-Pulse-Amplitude-Modulation (PAM4) for a net bit rate of 120 Gbit s$^{-1}$.

We designed a waveguide integrated PTE photodetector based on the split-gate geometry[52](Fig. 1).

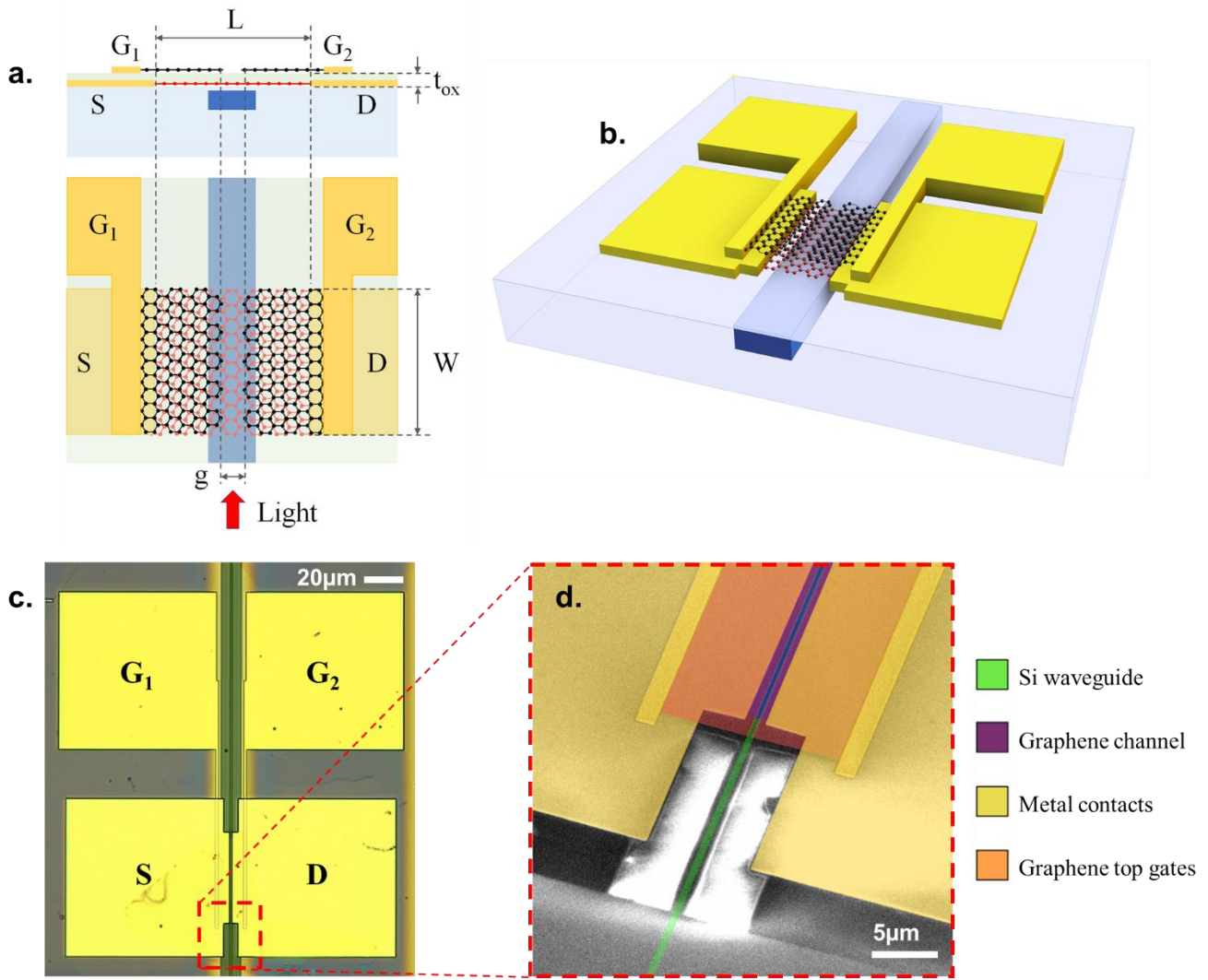

Figure 1. a. Cross section and top view of the designed PTE graphene photodetector. The active region is the first layer of graphene: the channel (red) of length L and width W, on top of the 480 nm x 220 nm Si waveguide (blue, light blue is SiO$_2$). The channel is accessed electrically by the source (S) and drain (D) metal contacts (yellow). A 100 nm thin SiN layer (light green) is used as dielectric spacer for the gates. Top gates are realized with a second layer of graphene (black) patterned in a split gate configuration with a gap (g) of 150 nm. Graphene gates are contacted through metal contacts (G$_1$ and G$_2$). b. 3D schematic of the proposed PTE graphene photodetector. c. Optical microscope picture of the fabricated devices: scale bar is 20 µm. d. False color SEM image of the input of the fabricated PTE graphene photodetector: green the Si waveguide, yellow the metal contacts, purple the active graphene channel (violet on the Si waveguide), orange the second graphene gates. Scale bar is 5 µm.

The photodetector has been integrated on top of a transverse electric (TE) silicon on insulator (SOI) waveguide designed for single mode operation at 1550 nm wavelength, with core cross section 480x220 nm (Fig.1a). The top cladding is thinned and planarized to a final thickness of about 20 nm on the top of the waveguide in order to maximize the interaction of the evanescent mode field with the graphene layer transferred on top of the silicon core (see Methods). The detector consists of a stack of two layers of graphene separated by a $t_{ox}$ = 100 nm thick SiN layer (Fig.1a). The photo-conversion occurs in the first single layer graphene where optical absorption of light generates a gradient of HCs' temperature (see Supplementary). The second layer of graphene is used to realize two split gates ($G_1$ and $G_2$) separated by a 150 nm large gap (g in Fig.1a), used to electrostatically induce a step change of the Seebeck coefficient in the first graphene layer. We refer to the distance between the source (S) and drain (D) metal contacts as the channel length L and to the width of the first graphene as the channel width W (Fig.1a). The generated photovoltage can be approximated as[49]:

$$V_{ph} = (S_1 - S_2)\Delta T \qquad (1)$$

where $S_1$ and $S_2$ are the Seebeck coefficients of graphene below gate 1 and gate 2 (fig. 1(a)), $\Delta T = \overline{T_{HC}} - T_0$ is the difference between of the HCs' temperature averaged over the graphene channel ($\overline{T_{HC}}$), which is proportional to the absorbed optical intensity, and the graphene lattice temperature ($T_0$). The detailed model for the calculation of the photovoltage is in the Supplementary Material. We remark that the magnitude of the Seebeck coefficient depends on graphene electronic properties, e.g. mobility and residual carrier concentration at the charge neutrality point, n* [49,53,54]. High quality graphene, i.e. having high mobility and low n*, exhibits higher Seebeck coefficient. E.g., experimental values <20 µV K$^{-1}$ for polycrystalline CVD graphene (grain size <5 µm, mobility <1000 cm$^2$ V$^{-1}$ s$^{-1}$), have been reported[55], while high quality exfoliated graphene on hexagonal Boron Nitride (hBN) substrate (mobility >10000 cm$^2$ V$^{-1}$ s$^{-1}$) has been demonstrated to exhibit S = 183 µV K$^{-1}$ [56]. In this work, we used a CVD grown single crystal graphene with mobility as high as 130000 cm$^2$ V$^{-1}$ s$^{-1}$ at room temperature when transferred on hBN substrates[57,58]. A single crystal graphene channel is important not only for the high Seebeck coefficient, but also for the absence of grain boundaries that may be responsible for non-uniform control of the Seebeck coefficient along the graphene channel[59].

The device geometry of Fig. 1 has been optimized to maximize the photovoltage while matching the impedance between the device and the read-out electronics, e.g. a digital sampling oscilloscope (DSO) or a discrete radio frequency (RF) voltage amplifier (See Supplementary).

We fabricated the device on a standard SOI platform with 220 nm thick Si overlayer and 2 μm thick buried oxide (BOX) (see Methods). We used high quality single crystal graphene grown by chemical vapor deposition (CVD) and transferred on the Si waveguides by a semi-dry transfer technique demonstrated previously[57]. After the transfer and patterning of the first layer of graphene, a stack of Nickel (Ni) and Gold (Au) was used to realize the source and drain metal contacts. Then, we transferred single-layer of hBN to protect the first layer graphene from the subsequent plasma-enhanced chemical vapor deposition (PECVD) of 100 nm of SiN used as gate dielectric. The second graphene layer was grown, transferred and patterned using the same procedures utilized for the first layer. Top gate contacts were deposited using the same process as for the first graphene channel. We used Raman spectroscopy[60] to characterize the two graphene crystals (see Methods).

We have experimentally characterized the static behavior of the PTE photodetector by mapping the photovoltage (Fig. 2a) and the resistance (Fig. 2b) as a function of the applied gate voltages. We coupled the light of a continuous wave (CW) laser source at 1550 nm wavelength into the SOI waveguide by means of a single mode optical fiber and a single polarization grating coupler (see Methods). We swept the bias applied to both the gate electrodes with respect to the channel source electrode between 0 V and -10 V and measured the photovoltage imposing zero current between the source and drain electrodes (see Methods). The device resistance map (fig. 2b) has been obtained in a similar way measuring the resistance of the detector for each couple of gate voltages in the absence of the optical excitation.

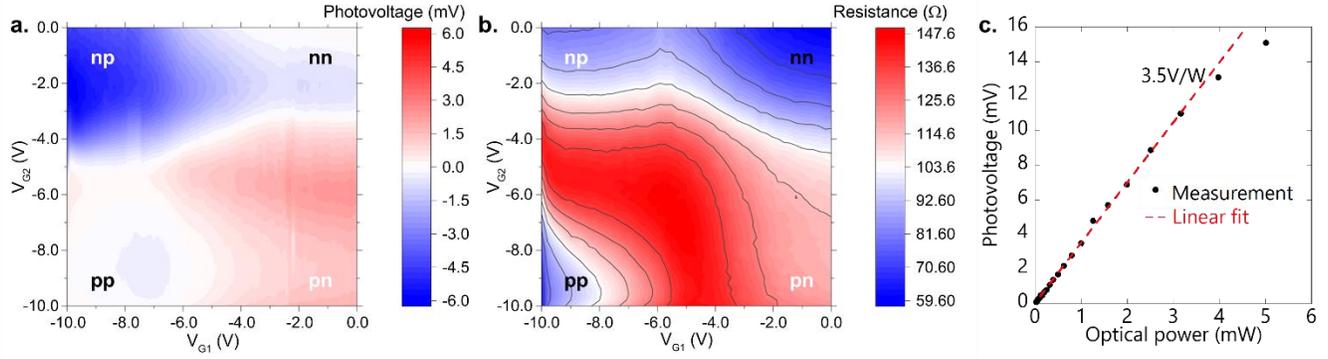

Figure 2. a. Measured photovoltage map of the fabricated PTE graphene photodetector (L = 1.5 µm and W = 50 µm) as a function of the gate voltages $V_{G1}$ and $V_{G2}$. Optical input was ~1.65 mW at the detector input facet. b. Measured resistance map of the fabricated PTE graphene photodetector (L = 1.5 µm and W = 50 µm) as a function of the gate voltages $V_{G1}$ and $V_{G2}$. c. Photovoltage as a function of the optical power at the detector input facet. Black dots are experimental values, the red dashed line represents the linear fit.

The photovoltage map clearly shows different regions where the photovoltage changes its sign, which is the signature of the dominance of PTE over other photoconversion mechanisms like photovoltaics at zero bias[50]. We show a photovoltage map having a maximum absolute value of about 6 mV for an optical input power at the detector equal to 1.7 mW (see Methods), which results in a voltage responsivity of about 3.5 V W$^{-1}$ for $V_{G1}$ = -1 V and $V_{G2}$ = -8 V. The asymmetric photovoltage map may be explained by an asymmetry in the energy dependence of the carrier relaxation time in graphene between the n- and p-type transport[56]. This behavior leads to an asymmetric Seebeck coefficient profile[56] with respect to the charge neutrality point. The device resistance in proximity of the maximum responsivity is about 90 Ω, which is higher than the designed 50 Ω. The difference can be ascribed to deviations in the sheet and contact resistances of the fabricated device with respect to the design. We measured the linearity of the photovoltage as a function of the input optical power (Fig. 2c) (see Methods). The photoresponse is linear up to 3 mW, with the slope corresponding to a voltage responsivity of 3.5 V W$^{-1}$. The low responsivity is probably due to the first graphene layer whose quality is deteriorated by the PECVD process deposition of the 100 nm thick SiN (see Raman characterization in Methods). Although the first graphene is protected by hBN to preserve the material quality, the process needs further improvements. An improved encapsulation[61] would give results closer to what is predicted by our simulation (see Supplementary). Moreover, the responsivity may be further increased by adopting optical absorption enhancement as demonstrated in ref.[20] and ref.[22].

We used the fabricated PTE photodetector to detect high baud rate optical data streams. In a first set of experiments, we used the device to detect a Non-Return-to-Zero (NRZ) On-off-Keying (OOK) optically modulated signal at 28 Gbit s$^{-1}$ (see Methods). PTE operation does not require an applied voltage between the source and drain electrode (bias free). We used a bias-tee to monitor the DC component of the photovoltage and to optimize the responsivity (see Methods). We found the maximum responsivity (~3V W$^{-1}$) for gate voltages close to the ones of the maximum $R_V$ in the DC map, i.e. $V_{G1} = -1$ V and $V_{G2} = -8$ V.

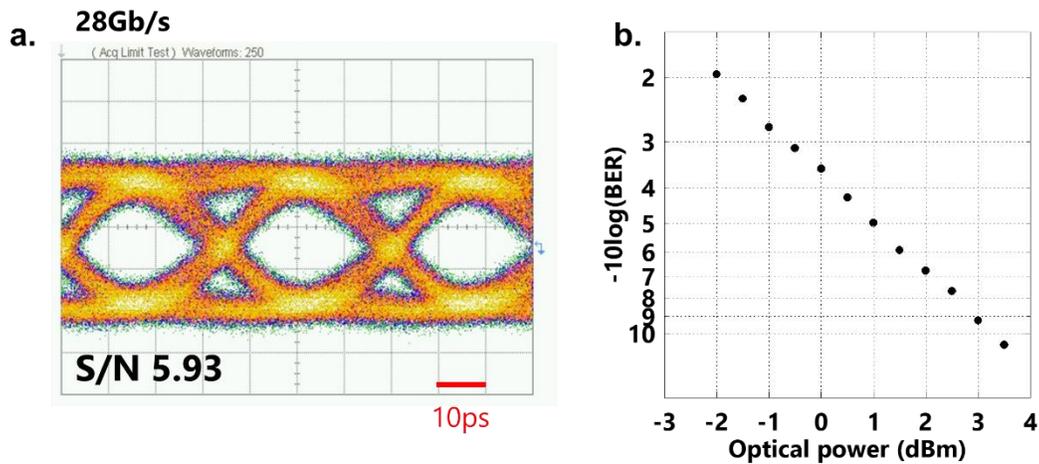

Figure 3. a. Eye diagram collected for a 28 Gbit s$^{-1}$ NRZ OOK data stream at the input of the PTE graphene photodetector. The optical power at the input facet of the photodetector was 2 mW. b. BER curve as a function of the optical power at the input facet of the detector. In both measurements, an electrical RF amplifier with 23 dB gain and 35 GHz bandwidth has been used.

A 50 Ω matched commercial RF amplifier with 23 dB gain and 35 GHz bandwidth has been used to amplify the electrical signal generated by the photodetector. Fig. 3 shows the collected eye diagram and corresponding Bit Error Ratio (BER) curve as a function of the optical power at the detector input (see Methods). A BER equal to 1.8x10$^{-11}$ has been obtained for an input optical power of 3.5 dBm in back-to-back configuration, i.e. by connecting through a short fiber (few meters) the detector to the transmitter.

In order to test our graphene detector at higher baud rates, we performed a second set of experiments involving a digital-to-analog converter (DAC) based optical transmitter with digital pre-emphasis. This is used to compensate for DAC and driver losses and for the limited bandwidth of the LiNbO$_3$ Mach Zehnder modulator (MZM) used to modulate the optical CW signal (see Methods). After back-to-back transmission the output voltage of the detector is amplified using a commercial amplifier with 22 dB gain and 65 GHz bandwidth. A real time oscilloscope and offline digital signal processing (DSP) is used for data acquisition, signal recovery and performance assessment. A block diagram of the setup is shown in fig. 4.

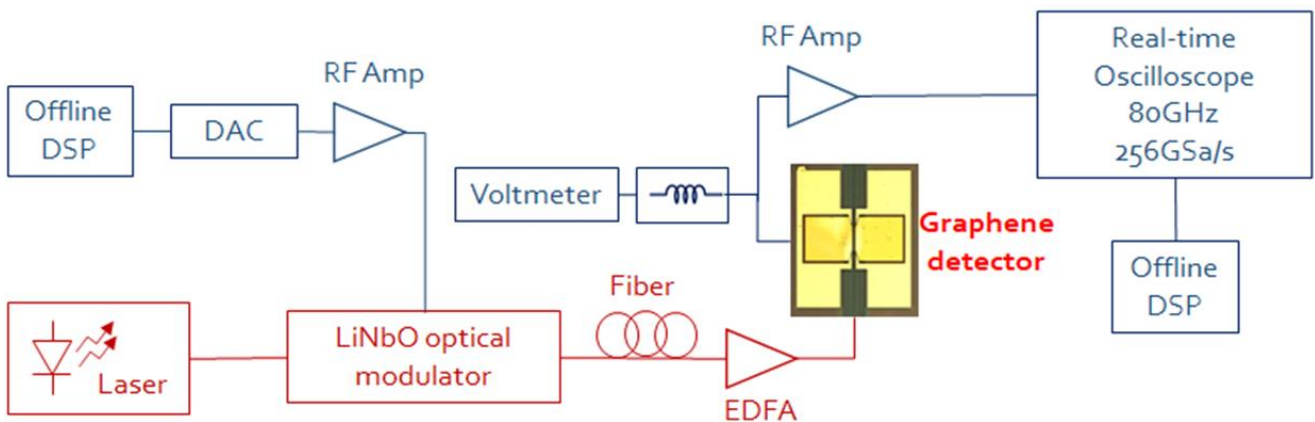

Figure 4. High speed optical testbed with a DAC based transmitter and an ADC based receiver.

We tested our device at 60 GBaud using an NRZ OOK and a four level Pulse Amplitude Modulation (PAM4) format, and at 105 GBaud NRZ OOK. The collected eye diagrams and the corresponding histograms of the signal amplitude at the optimum sampling point are shown in fig. 5a-d.

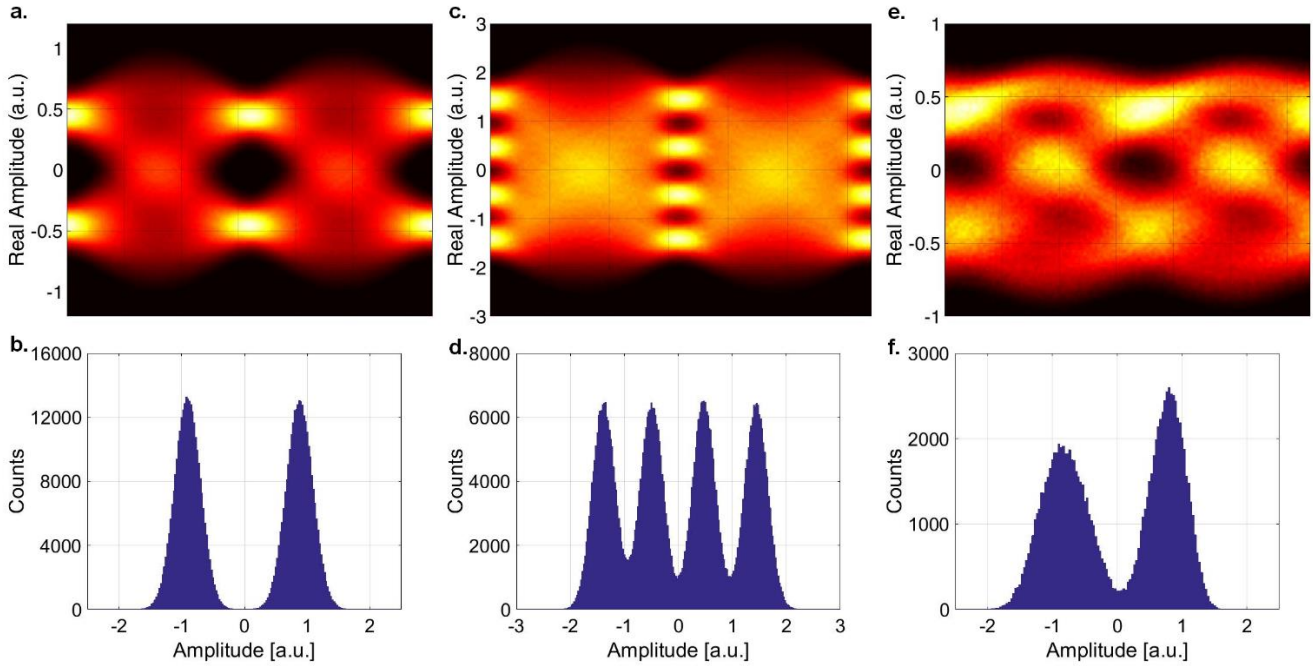

Figure 5. Eye diagrams collected for a 60 GBaud NRZ OOK (a), 60 GBaud PAM4 (c) and 105 GBaud NRZ OOK (e) data stream at the input of the detector. Histograms of the signal amplitude collected at the optimum sampling point for a 60 GBaud NRZ OOK (b), 60 GBaud PAM4 (d) and 105 GBaud NRZ OOK (f) data stream at the input of the detector.

The evaluated SNR values of the collected data are 12.5 dB for the NRZ data transmission and 13.7 dB for the PAM4. The corresponding BER is $8 \times 10^{-5}$ (NRZ) and $1.1 \times 10^{-2}$ (PAM4). In order to show the high bandwidth of the device, we repeated the experiment up to 105 GBaud NRZ data transmission. The collected eye diagram (Fig. 5e) exhibits SNR = 8.4dB with a resulting BER of $8 \times 10^{-3}$. The associated signal amplitude at the optimum sampling point (Fig. 5f) shows an asymmetry towards the zero level because of a non-optimized signal at the transmitter side.

In fig. 6a-c we show the electrical spectra of the received signals of fig. 5 obtained through a Fast Fourier Transform (FFT) of the real time signal after the equalization filter (see Methods).

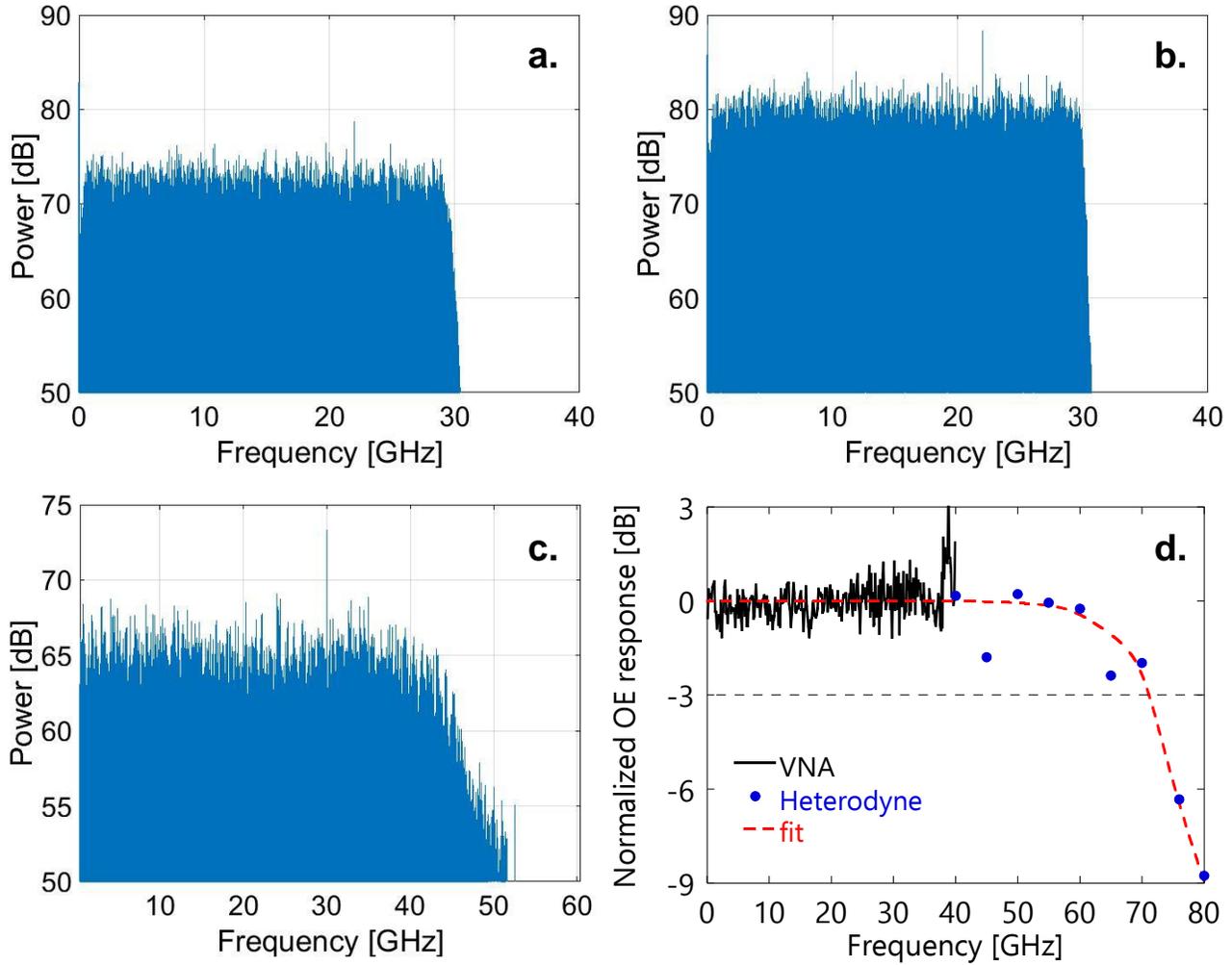

Figure 6. Electrical spectra of the received signals: 60 GBaud NRZ OOK (a), 60 GBaud PAM4 (b) and 105 GBaud NRZ OOK (c). d. Optical to electrical bandwidth of the received signal: black curve is the measurement as acquired from the VNA, blue dots are the measured response as obtained by the heterodyne setup with laser beating, the red dashed line is the interpolation curve.

The spectra are completely flat in the range of frequencies of the input electrical signal as generated at the transmitter side without any ripples or a drop at high frequencies. This demonstrates the high speed of operation of our detectors, as already demonstrated for a similar device in our previous work where we showed flat response up to 40GHz[52]. The response shown in Fig 6c. is flat up to 45 GHz, beyond this we see a 10 dB drop of the power at signals Nyquist frequency. In order to understand if it is related to the photodetector, we investigated the flatness of the frequency response by means of a Vector Network Analyzer (VNA, see methods) up to 40 GHz, which is shown in fig. 6d. The peak close to 40 GHz is ascribed to the electrical setup. The frequency response above 40 GHz up to 80 GHz was then measured by implementing a heterodyne setup superimposing two CW tunable lasers having frequencies $f_1$ and $f_2$, which are coupled through a 3dB fiber coupler and interfere causing a response at the beating frequency $f_{beating}=f_2-f_1$. We set $f_1$ constant and varied $f_2$ in order to have $f_{beating}$ in the range 40 GHz-80 GHz at steps of 5 GHz. We obtained the frequency response by numerically computing the FFT of the acquired real time traces, and extracting the peak intensity of the signal in the frequency domain. Results are shown as blue dots in fig. 6d. The frequency response is flat up to 60 GHz and then start decreasing because of the roll-off of the used RF amplifier (SHF804B specified flat up to 60 GHz). From the extracted frequency response, we can conclude that the drop shown in fig. 6c is not arising from the graphene photodetector. The origin of this roll-off will be investigated with further experiments.

In conclusion, we demonstrated a waveguide integrated graphene photodetector based on the PTE effect operating at zero bias without dark current. The device exhibits a responsivity of about 3.5 V W$^{-1}$ with a flat frequency response up to 65 GHz. In this work, for the first time to the best of our knowledge, we demonstrated the detection of an optical data transmission by using different modulation formats at baud rates up to 105 GBaud, by using the PTE effect in unbiased operation. This work demonstrates the feasibility of using PTE based graphene detectors as optical receivers and paves the way for the realization of versatile and highly efficient optical graphene transceivers.

**Methods**

Fabrication

We used electron beam lithography (EBL) to realize single mode transverse electric (TE) polarized Si photonic waveguides with cross section 480x220 nmon a standard SOI substrate with 220 nm thick Si overlayer and 2 μm thick BOX. The waveguide is provided with single polarization grating couplers designed to couple the light to the fundamental TE mode of the Si waveguide. The waveguides were then covered with a with a thin layer of tetraethyl orthosilicate (TEOS) and a thick boron-phosphorus TEOS (BPTEOS) cladding which was next thinned down to a final thickness of ~25 nm on top of the waveguide. Graphene was grown on copper (Cu) foils by deterministic seeded growth by chemical vapor deposition (CVD)[57]. We used Chromium (Cr) nucleation seeds the Cu foils and patterned by optical lithography. Regular arrays of graphene crystals were grown at 1060 °C with background pressure of 25 mbar using a 4" BM Pro cold-wall reactor. Sample enclosure and Argon (Ar) annealing were used to control the nucleation density[57]. After growth, a PMMA layer was spin-coated on the Cu foil and an adhesive frame was attached to the perimeter of the sample. Electrochemical delamination of the graphene crystals was used to separate graphene from the Cu growth substrate by applying a voltage of −2.4 V with respect to a Cu counter-electrode in a 1M NaOH solution. The delaminated graphene array on the PMMA membrane was thoroughly rinsed in deionized water. We then aligned the array precisely to the target waveguides using a micrometric 4-axis stage and laminated it at 90 °C. The sample was heated at 105 °C for 5 minutes to improve the adhesion. The PMMA support was then removed in acetone. EBL and reactive ion etching (RIE) (5 sccm oxygen and 80 sccm argon, 35 W power) were used to pattern graphene by using PMMA as an etch mask. After the transfer and patterning of the first graphene layer, metal has been thermally evaporated to realize top contacts. We used a stack 7 nm of Ni and 60 nm of Au, followed by lift-off in acetone. The fabricated graphene channel was next coated with a protective layer of a commercially available large-area polycrystalline single-layer hBN provided by Graphene Laboratories Inc. hBN was used to provide protection to the first layer graphene with respect to the plasma used during the subsequent PECVD (350 °C) of the 100 nm SiN dielectric film. The second graphene layer used as gates was grown, transferred and patterned with the same process described for the first layer. Finally, metal contacts to the graphene gates have been realized with the same thermal evaporation of Ni and Au.

Raman characterization

Raman spectroscopy has been used to characterize the quality of the SLGs used for photodetectors fabrication. All the Raman spectra have been collected in a Renishaw InVia spectrometer, equipped with a 532 nm excitation laser, at laser power ~1 mW and acquisition time of 4 s. The measurements have been performed for each fabrication step (See Supporting Information). Starting from pure transferred graphene, the Raman spectra has a 2D peak with a single Lorentzian shape and FWHM(2D) ~25 cm$^{-1}$, showing a low level of sub-µm strain fluctuations (and thus relatively high carrier mobility)[62], comparable to exfoliated graphene on SiO2. Due to subsequent hBN transfer and SiN deposition an increase of the FWHM(2D) peak is reported, obtaining ~25.5 cm$^{-1}$ and ~32.5 cm$^{-1}$, respectively, indicating an increase in strain fluctuations and thus lower carrier mobility.

DC Measurements

Static characterization of the graphene photodetector has been performed by using two electrical source-meters: the first was used to sweep the gate voltages with two independent channels, the second was used to set an open circuit (zero current) between source and drain and to read the generated photovoltage. A ground-signal (GS) probe was used to probe the source and drain, while DC needle probes were used to contact the gates. A CW laser source was amplified by an Erbium-doped fiber amplifier (EDFA) and coupled to the chip with a single mode optical fiber coupled to the input grating coupler. A fiber-based polarization controller was used to match the required polarization at the input grating coupler. After the input grating coupler, we used an integrated 3 dB splitter: one output branch of the splitter was coupled to the PTE photodetector, while the other branch was used to monitor the optical power through an output grating coupler. After the PTE photodetector, another grating coupler was used to monitor the optical power at the device output. In this way we could perform a precise characterization of the insertion loss of the device and properly evaluate the optical power at the device input. We estimated a total insertion loss of -13 dB due to the grating coupler efficiency (-5 dB), the 3dB splitter (-3 dB) and propagation loss in the access waveguide (-5 dB). This last is due to particle (polymers and metal) deposition on the access waveguide which is not protected with cladding. Optical power dependent measurements were done by using a fiber-based variable optical attenuator (VOA) after the EDFA.

Broad band measurement

The first set of measurements was performed by using a commercial pattern generator and BER tester at 28 Gbits$^{-1}$. A LiNbO$_3$ MZM having 30 GHz bandwidth was used to modulate the optical signal. A variable optical attenuator (VOA) was used to perform the BER measurements as a function of the input optical power. A 50Ω matched commercial RF amplifier with 23 dB gain and 35 GHz bandwidth was used to bring the electrically generated signal to a level readable by a digital sampling oscilloscope (DSO) and by the BER tester.

The chip level Graphene photodetector was interconnected on a probe station with the same fiber coupling scheme described above. For RF interconnection we used a 67 ground-signal (GS) probe to access the electrical signal generated by the photodetector. A 67 GHz bandwidth bias-tee was used to impose zero DC current to the device.

We used the same MZM of the previous setup for the measurement of the frequency response. Port A of the VNA was used to drive the MZM and the second port to measure the scattering parameter S21 at the output of the graphene photodetector. We used a commercial 70 GHz bandwidth photodetector to measure the response of the setup, i.e. modulator, cables, probes and connectors. The electrical amplifier was not used for the frequency response measurement.

The second set of measurements was performed at higher speed. In the transmitter we used an 88 GSa $s^{-1}$ DAC to generate the PAM data. Pre-calculated data were sent out periodically. In the transmitter digital signal processor, we applied raised cosine pulse shaping and in addition we compensated for the DAC, driver amplifier and modulator low pass characteristics. The electrical signal is modulated onto a 1550 nm optical carrier supplied by an external cavity laser, the modulator was biased at quadrature. The optical signal was amplified using an EDFA before launching onto the photodetector. The electrical response of the device is amplified using a 65 GHz bandwidth amplifier having 22 dB gain. Next, we digitized the data in a real time oscilloscope at 33 GHz bandwidth and stored the data for subsequent digital signal processing (DSP). The offline DSP incorporates resampling to 2 Sample/symbol to simplify filter implementation. Next, we applied the channel compensation using a linear feed forward equalizer, where the filter taps where adapted respectively. Finally, we assessed the performance by counting the bit errors to determine the bit error ratio (BER) as well as the signal to noise ratio (SNR).

In a third experiment we further increased the speed by replacing the DAC and ADC. In the transmitter we applied a 120 GSa $s^{-1}$ DAC and generated NRZ data at 105 GBaud. The data where amplified in a 65 GHz driver and modulated using LiNbO$_3$ modulator. In the receiver we digitized the data at 256 GSa s$^{-1}$ and 80 GHz bandwidth. Further details were identical to the second experiment.

graphene on SiC. 4–7 (2017).

46. Kikuchi, K. Fundamentals of coherent optical fiber communications. *J. Light. Technol.* **34**, 157–179 (2016).

47. Tielrooij, K. J. *et al.* Photoexcitation cascade and multiple hot-carrier generation in graphene. *Nat. Phys.* **9**, 248–252 (2013).

48. Tielrooij, K. J. *et al.* Generation of photovoltage in graphene on a femtosecond timescale through efficient carrier heating. *Nat. Nanotechnol.* **10**, 437–443 (2015).

49. Gabor, N. M. *et al.* Hot carrier-assisted intrinsic photoresponse in graphene. *Science (80-. ).* **334**, 648–652 (2011).

50. Song, J. C. W. W., Rudner, M. S., Marcus, C. M. & Levitov, L. S. Hot carrier transport and photocurrent response in graphene. *Nano Lett.* **11**, 4688–4692 (2011).

51. Lemme, M. C. *et al.* Gate-activated photoresponse in a graphene p-n junction. *Nano Lett.* **11**, 4134–4137 (2011).

52. Marconi, S. *et al.* Waveguide Integrated CVD Graphene Photo-Thermo-Electric Detector with >40GHz Bandwidth. *2019 Conf. Lasers Electro-Optics, CLEO 2019 - Proc.* **1**, 4–5 (2019).

53. Hwang, E. H., Rossi, E. & Das Sarma, S. Theory of thermopower in two-dimensional graphene. *Phys. Rev. B - Condens. Matter Mater. Phys.* **80**, 1–5 (2009).

54. Sierra, J. F., Neumann, I., Costache, M. V. & Valenzuela, S. O. Hot-Carrier Seebeck Effect: Diffusion and Remote Detection of Hot Carriers in Graphene. *Nano Lett.* **15**, 4000–4005 (2015).

55. Lim, G. *et al.* Enhanced thermoelectric conversion efficiency of CVD graphene with reduced grain sizes. *Nanomaterials* **8**, 1–9 (2018).

56. Duan, J. *et al.* High thermoelectricpower factor in graphene/hBN devices. *Proc. Natl. Acad. Sci. U. S. A.* **113**, 14272–14276 (2016).

**Acknowledgment**

We acknowledge Prof. A. C. Ferrari and Prof. F. H. L. Koppens for fruitful discussion on photo-thermal effect in graphene photodetectors.

**Funding**

This work was supported by the European Union's Horizon 2020 research and innovation programme, GrapheneCore2 under grant 785219

**Author contribution**

All the authors conceived the detector within the Graphene Flagship project. S.M., V.S., M.R. did the design of the detector. V.M., C.C. prepared the graphene crystals and performed the graphene characterization. M.A.G., V.M., S.S., S.T., performed the device fabrication. S.M., V.S., M.A.G., A.M., F.B., did the device characterization and system testing. S.M., V.S., F.B., P.G., W.T., M.R. did the analysis of the results.

# Supplementary Material to

Photo Thermal Effect Graphene Detector Featuring 105 Gbit s[-1] NRZ and 120 Gbit s[-1] PAM4 Direct Detection

S. Marconi[1], M. A. Giambra[2], A. Montanaro[2], V. Mišeikis[3,4], S. Soresi[2,5], S. Tirelli[2,5], P. Galli[6], F. Buchali[7], W. Templ[7], C. Coletti[3,4], V. Sorianello[2] and M. Romagnoli[2,*]

[1]Tecip Institute – Scuola Superiore Sant'Anna, Via G. Moruzzi 1, 56124 Pisa, Italy

[2]Photonic Networks and Technologies Lab – CNIT, Via G. Moruzzi 1, 56124 Pisa, Italy

[3]Center for Nanotechnology Innovation @NEST - Istituto Italiano di Tecnologia, Piazza San Silvestro 12, I-56127 Pisa, Italy

[4]Graphene Labs, Istituto Italiano di Tecnologia, Via Morego 30, 16163 Genova, Italy

[5]Fondazione INPHOTEC, Via G. Moruzzi 1, 56124 Pisa, Italy

[6]Nokia Solutions and Networks Italia, via Energy Park 14, 20871 Vimercate, Italy

[7]Nokia Bell Labs, Lorenzstr. 10, 70435 Stuttgart, Germany

*corresponding author: marco.romagnoli@cnit.it

## S.1 Mathematical model of PTE based waveguide integrated photodetector

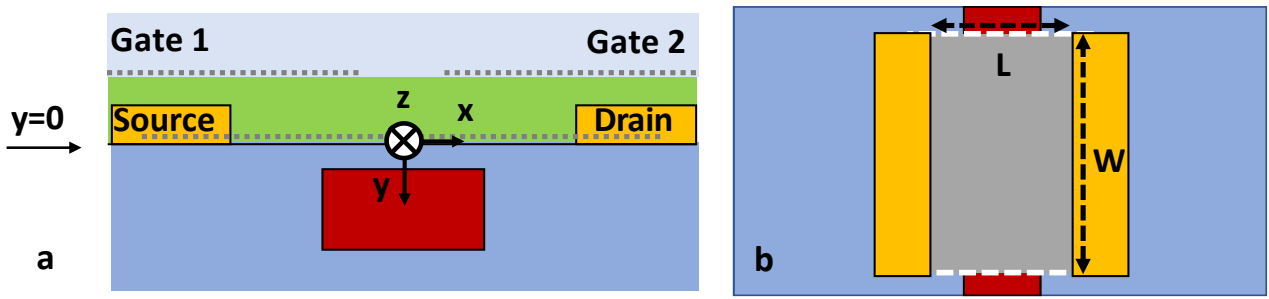

*Fig.S1. a) Cross section of the realized device. b) In plane cut of the active graphene layer, (x,z) plane at y=0.*

The mathematical model used in the main text to compute the photovoltage and the voltage responsivity $R_V$ is summarized by the set of equations s1.

$$\nabla \cdot (k_{HC} \nabla T_{HC}) + \frac{k_{HC}}{L_C^2}(T_{HC} - T_{lattice}) = \eta \frac{P_{in}(x)\exp\left(-\frac{z}{L_a}\right)}{L_a} - \vec{J}\cdot \nabla \Pi \quad \text{(s1.a)}$$

$$\vec{J} = \sigma(-\nabla V + S\nabla T_{HC}) \quad \text{(s1.b)}$$

$$\nabla \cdot \vec{J} = 0 \quad \text{(s1.c)}$$

All the parameters and variables are defined in the channel plane (active graphene layer, (x,z) plane in y=0, Fig.S1a-b). Eq. s1.a is the heat transport equation as defined in the work by Song et al.[1], which relates the absorbed optical power density to the hot carriers' (HCs) temperature. Eq. s1.b describes the current density $\vec{J}(x,z)$ in the graphene channel in the HCs regime accounting for photo-thermoelectric (PTE) contribution. Eq. s1.c is the continuity equation in stationary condition. The operator $\nabla = \left(\frac{\partial}{\partial x}, \frac{\partial}{\partial z}\right)$ is defined in the plane y=0 and acts on the x and z coordinates. $T_{HC}(x,z)$ and $T_{lattice}$ are the HCs and lattice temperature while $V(x,z)$ is the electrical potential in the active graphene channel. S and $k_{HC}$ are the Seebeck coefficient and the HCs thermal conductivity in graphene, respectively. The Peltier term is computed as $\Pi=ST$ [1,2]. The term $(1/L_C^2)(T_{HC} - T_{lattice})$ is the term associated to the cooling of HCs and $L_C$ is the parameter known as cooling lenght[1]. We assumed $L_C$ = 140 nm as reported in ref.[3] for polycrystalline graphene. $\eta P_{in}(x)\exp(-z/L_a)/L_a$ is the source term, i.e., the optical power density absorbed at coordinate z (along the light propagation direction) in a point of the graphene channel (y=0) and therefore delivered to the hot carriers system. $L_a$ is the absorption length of the optical mode and $\eta$ is the fraction of the absorbed optical power which is absorbed in the channel layer and not by the graphene gates. The parameters $L_a, \eta$, as well as $P_{in}(x)$, have been obtained by means of the modal analysis of the structure in Fig.S1.a, performed using a commercial mode solver, and shown in the section S.3.

We used $T_{HC} = T_{lattice}$ as boundary condition in x = ± L/2, V = 0 V in x = -L/2 and V = $V_d$ in x = L/2. The photocurrent is computed as

$$I_{ph} = \int_0^W J_x\left(x = \frac{L}{2}, z\right) dz \qquad (s2)$$

and the photovoltage $V_{ph}$ is obtained as the drain voltage when $I_{ph}$ is null.

## S.2 Conductivity model

The Seebeck coefficient S and the thermal conductivity $k_{HC}$ have been computed by using the Mott's formula[1,4] ($S = -\frac{\pi^2 k_B^2 T}{3e} \frac{1}{\sigma} \frac{d\sigma}{d\mu_c}$ where $k_B$ is the Boltzmann constant, e the electron charge and $\mu_C$ is the chemical potential in the graphene channel) and the Wiedemann-Franz law[1,5] ($k_{HC} = \sigma LT$, L is the Lorentz number). A conductivity model similar to the one proposed in Ref[6] has been used

$$\sigma = \sigma_{min}\sqrt{1 + \frac{n^2}{\Delta n^2}} \qquad (s3)$$

where n is the charge carrier concentration and $\Delta n$ is the magnitude of the carrier density fluctuations in graphene in the proximity of the charge neutrality point. By using the relation

$$n = \frac{1}{\hbar^2 v_F^2 \pi} \mu_C^2 \qquad (s4)$$

($\hbar$ is the reduced Planck's constant, $v_F$ is the Fermi velocity in graphene), we can write

$$\sigma = \sigma_{min}\sqrt{1 + \frac{\mu_C^4}{\Delta^4}} \qquad (s5.a)$$

$$\Delta = \hbar v_F \sqrt{\pi \Delta n} \qquad (s5.b)$$

We use Δ = 100 meV (Δn = 7x10$^{11}$cm$^{-2}$) and a $\sigma_{min}$ equal to 5x10$^{-4}$S (2000Ω/sq. at the charge neutrality point and 1000 Ω/sq. in proximity of the maximum responsivity point). Those values are consistent with the ones estimated in a previous work[7]. We assumed the chemical potential to be constant in the regions under the gate electrodes and to smoothly vary in the 150 nm gap region. The photovoltage map in Fig.2.d of the main text has been simulated by sweeping the chemical potential on both sides of the junction. The corresponding gate voltage has been obtained by using the relation $en \approx C_{gate}(V_{gate} - V_{CNP})$ and $C_{gate} = \varepsilon_0 \varepsilon_{SiN}/t_{SiN}$ is the gate capacitance per unit area ($\varepsilon_0$ is the vacuum permittivity, $\varepsilon_{SiN} = 6$ is the relative dielectric constant used for SiN and $t_{SiN}$ is gate dielectric thickness).

### S.3 Optical simulation and heat source term

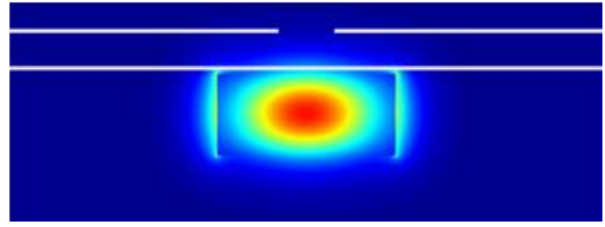

*Fig.S2. Electric field intensity profile of the fundamental TE mode sustained by the waveguide in Fig. S1.a (Si core 480x220 nm, λ=1550nm).*

The absorption length $L_a$, the power density $P_{in}(x)$ in the active graphene layer, the parameter η, and the minimum distance of the source-drain contacts from the waveguide to prevent losses and the optimal gate dielectric thickness are obtained from the simulation of the fundamental TE mode (Fig.S2) of the SOI waveguide with the detector stack (Fig.S1.a), at a wavelength equal to 1550 nm and propagating in the z-direction. The optical properties of the graphene layers have been modeled by using a surface conductivity model[8].

Optical power propagating in an absorbing medium (z-direction) undergoes to an exponential attenuation[9]

$$P_{opt}(z) = P(z=0)\exp(-\alpha z) = P(z=0)\exp(-z/L_a) \qquad (s6)$$

In eq. s6 $\alpha = 2\bar{\alpha}$ is the power attenuation constant, $\bar{\alpha}$ is the attenuation constant of the electric field amplitude and $L_a = 1/\alpha$ is the absorption length[9]. The amount of power absorbed between a point in z and z+dz in the active graphene layer, where dz is an infinitesimal distance, is, therefore,

$$-\frac{dP_{opt}}{dz} = \frac{1}{L_a}P(z=0)\exp\left(-\frac{z}{L_a}\right) \qquad (s7)$$

Since in the case of a waveguide integrated photodetector we have an optical mode, $L_a$ is obtained from the complex propagation constant. Due to the transversal spatial extension of the optical mode, the absorbed power density has a spatial profile along the x-coordinate that can be obtained from the z-component of the Poynting vector $P_z(x,z=0)$ in the (x,z) plane at y=0.

We assume that for an optical mode propagating in absorbing layers of infinite length, the mode optical power $P_{in}^*$ at the detector input is completely absorbed.

$$\int_0^{+\infty}\int_{-\infty}^{+\infty}\left(-\frac{dP_{opt}}{dz}\right)dxdz = \int_{-\infty}^{+\infty}\frac{1}{L_a}P_{in}(x)dx\int_0^{+\infty}\exp\left(-\frac{z}{L_a}\right)dz = P_{in}^* \qquad (s8)$$

From eq. s8 we obtain the following condition:

$$\int_{-\infty}^{+\infty}P_{in}(x)dx = \int_{-\infty}^{+\infty}A\frac{1}{2}Re(P_z(x,y=0,z=0))dx = P_{in}^* \qquad (s9)$$

(A is a normalization constant).

Part of the optical power is absorbed in the graphene gate electrodes and does not contribute to the photovoltage generation. The term η accounts for the fraction of the absorbed power which is effectively used in the photoconversion and is defined as:

$$\eta = \alpha_{active\ layer}/\alpha_{total} \qquad (s10)$$

We obtained $\alpha_{total}$ by simulating the detector with the full stack, including gates (Fig. S1a and S2) and $\alpha_{active\ layer}$ by removing the upper graphene layer. We assumed the presence of the graphene gates does not strongly modify the spatial profile of the optical mode as it may happen with metal gates, and the only parameter that changes in the two simulations is the power absorption.

The heat source term in eq. s1.a is, thus, written as

$$P_{absorbed}(x,z) = \eta \frac{P_{in}(x)\exp\left(-\frac{z}{L_a}\right)}{L_a} = \eta \frac{P_{in}^*}{\int_{-\infty}^{+\infty}\frac{1}{2}Re(P_z(x,y=0,z=0))dx} Re\left(\frac{1}{2}(P_z(x,y=0,z=0))\right)\frac{\exp\left(-\frac{z}{L_a}\right)}{L_a} \qquad (s11)$$

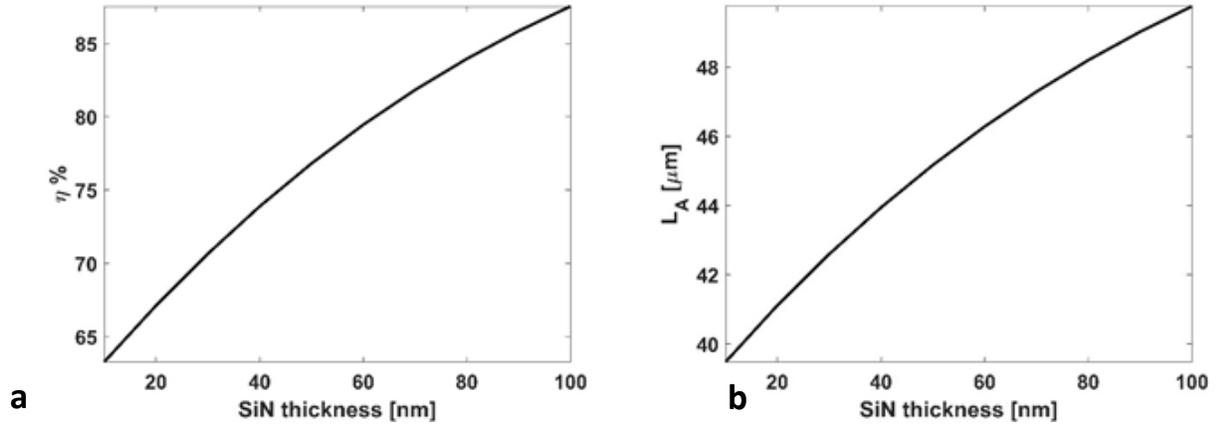

*Fig.S3. a) η% as a function of the SiN gate dielectric thickness. b) Absorption length $L_a$ as a function of the gate dielectric thickness*

The absorption length $L_a$ and η depend on the SiN thickness. For a 100nm thick SiN gate dielectric layer almost the 90% of the optical power is absorbed in the active graphene layer and the absorption length is about 50μm.

We avoided the use of a thicker dielectric layer because, given the low gate capacitance, a very large gate voltage would have been otherwise required to set the maximum responsivity.

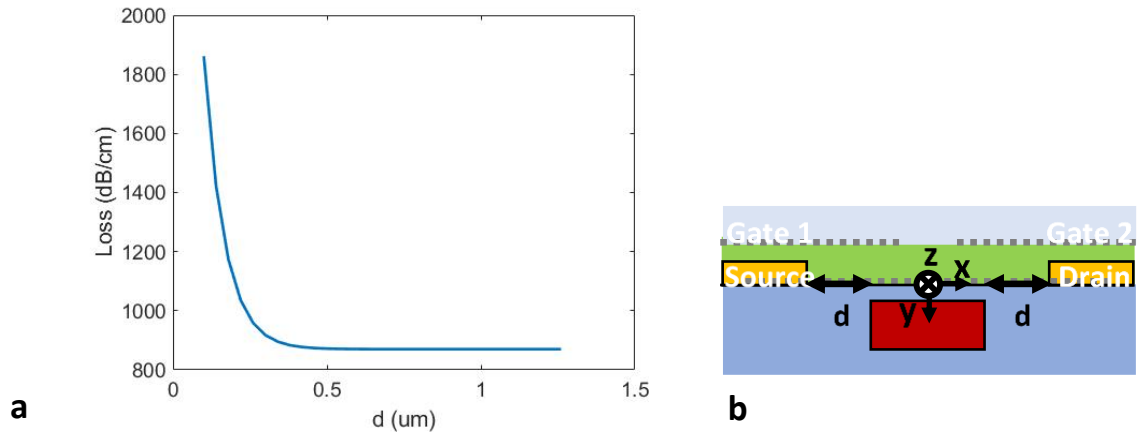

*Fig. S4. a) Losses of the optical mode as a function of the distance between metal contacts and waveguide. b.) Cross section of the device with contacts.*

We simulated the structure also in presence of the source/drain metal electrodes and we varied the distance d (Fig.S4.b) between the metal contacts and the waveguide. A distance d = 500 nm is sufficient to prevent extra losses introduced by metals (Fig.S4.a). Considering the width of the Si waveguide (480 nm) we can use a channel length L equal to 1.5µm (see main text).

## S.4 Voltage responsivity drop for device having large width W

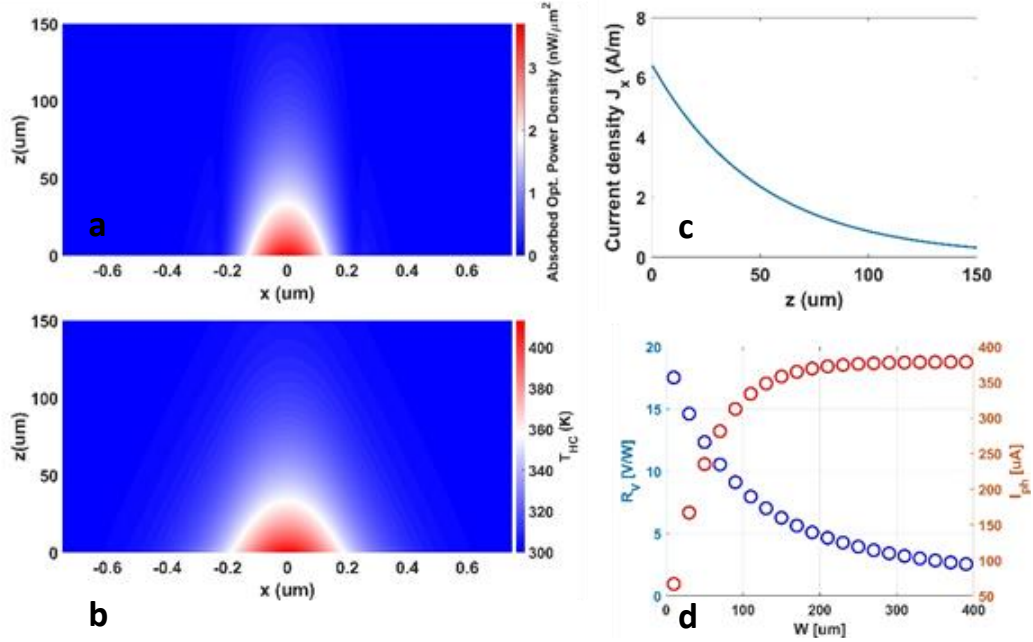

*Fig. S5 a) Heat source as defined in eq.s11 for a device having a channel length L=1.5μm, a channel width W=150μm, absorption length $L_a$≈50μm and $P_{in}$=1mW. b) Spatial profile of $T_{HC}$ (numerical solution of eq.s1) associated to the heat source in Fig.S5.a. c) x-component of the current density (numerical solution of eq. s1) along the drain contact surface (x=L/2), heat source in Fig.S5a . d) Photocurrent $I_{ph}$ as defined in eq.s2 obtained from the current density in Fig.S5.c.*

The density of absorbed optical power exponentially decreases (exp(-z/$L_a$)) along the propagation direction z (eq. s11, Fig.S5.a). The HCs temperature undergoes to the same exponential damping (Fig.S5.b) as the current density (Fig.S5.c). For z>>$L_a$ ($L_a$≈50μm in Fig.S5) the photocurrent density $J_x$ weakly contributes to the total photocurrent $I_{ph} = \int_0^W J_x\left(x = \frac{L}{2}, z\right) dz$. The photovoltage $V_{ph}$ and the voltage responsivity $R_V=V_{ph}/P_{in}$ decrease with the channel width W(Fig.S5.d, red curve). Indeed, $V_{ph}$ (the open circuit voltage of the current-voltage characteristic under illumination) and $I_{ph}$ (the short circuit current) are linked through the relation $V_{ph}=RI_{ph}$[10]. For large channel width (W>>$L_a$) the photocurrent saturates (Fig.S5d, blue curve) and the device resistance is reduced, therefore, the photovoltage drops (FigS5d, red curve).

## S.5 Series Impedance of the photodetector

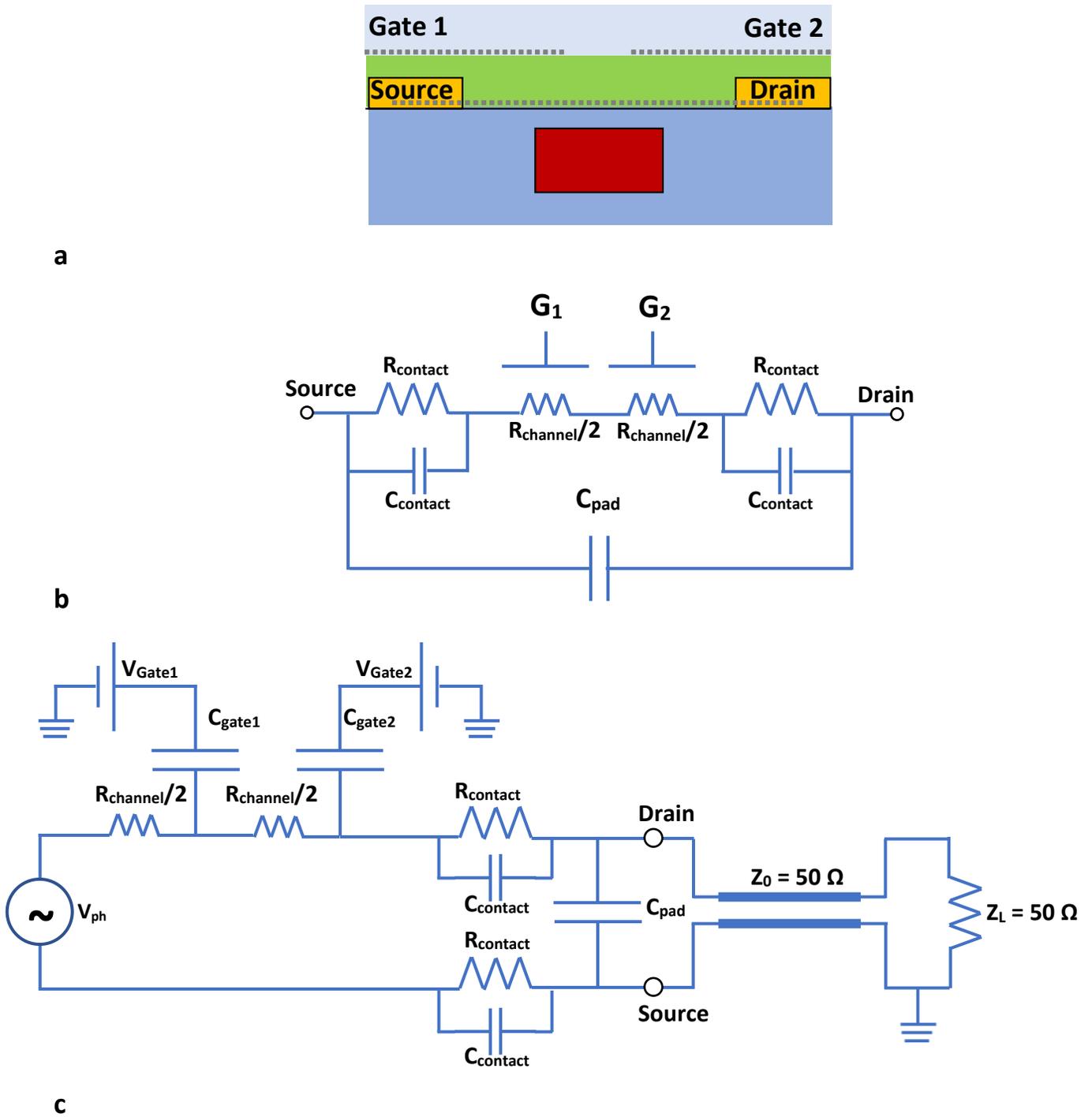

*Fig. S6 a) Cross section of the device and b) schematic representation of the resistance and capacitance distribution in the graphene channel under the gate electrodes. c) Equivalent electric circuit of the PTE photodetector when connected to a 50 Ω load through a lossless transmission line (characteristic impedance $Z_0$ = 50 Ω).*

Fig.S6c shows an equivalent electrical circuit model of the graphene photodetector. We modelled the photodetector as a voltage source with a series impedance. The capacitances due to the presence of the gates may affect the frequency response of the photodetector and introduce a cut-off frequency reducing the optoelectronic bandwidth. The electrical modeling is complicated by the fact that the lowest gate capacitor plate is a portion of the graphene active channel (Fig. S6.a-b). We model the gate capacitive effect as two capacitors placed in the half and at the end of the graphene channel (Fig. S6.c).

$$R_{contact} = \frac{R_C}{W} \tag{s12.a}$$

$$R_{channel} = \frac{R_{ch}L}{2W} \tag{s12.b}$$

$$C_{gate,1,2} = \frac{\varepsilon_0 \varepsilon_{SiN}}{t_{SiN}} \frac{L}{2} W \tag{s12.c}$$

where $R_c$=500 Ω µm and $R_{ch}$ = 1000 Ω/sq. (as estimated in a previous work[7]), LW/2 is the area of the gate capacitor plate. The other parameters like the contact capacitance $C_C$ and the pad capacitance $C_{pad}$ are assumed to be similar to the ones reported in ref[11].

| $R_{channel}$ | $R_{contact}$ | $C_{gate,1,2}$ | $C_{contact}$ | $C_{pad}$ |
|---|---|---|---|---|
| 30 Ω | 10 Ω | 20 fF | 0.12 pF | 1.2 fF |

*Table I. Electrical parameters of the circuit in Fig. S6.c computed for a device geometry L = 1.5µm and W = 50 µm or taken from ref.[11]*

We simulated the frequency response of the circuit in Fig. S6.c. (Fig. S7) by using the parameters in Table I. The frequency response is almost flat for f < 100 GHz (attenuation of 0.7 dB at f = 100 GHz with respect to the low frequency value). In this frequency range the device behavior is mainly determined by the resistive component of its impedance which limits the power transfer to the 50 Ω load. Since our experimental set-up is limited to f < 100 GHz, we optimized the PTE photodetector by considering only the channel and the contact resistance and neglecting the capacitive effects (simplified circuit in Fig. S8).

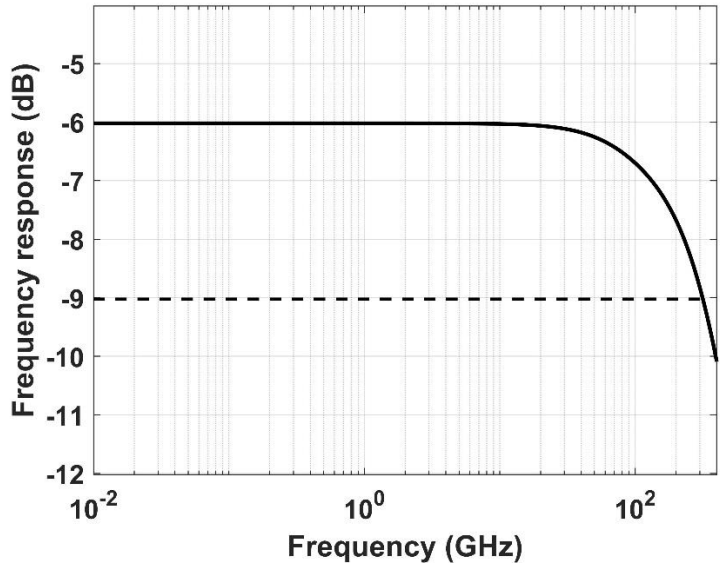

*Fig.S7. Frequency response (output on $Z_L$) of the circuit in Fig.S6c using the parameters in Table I.*

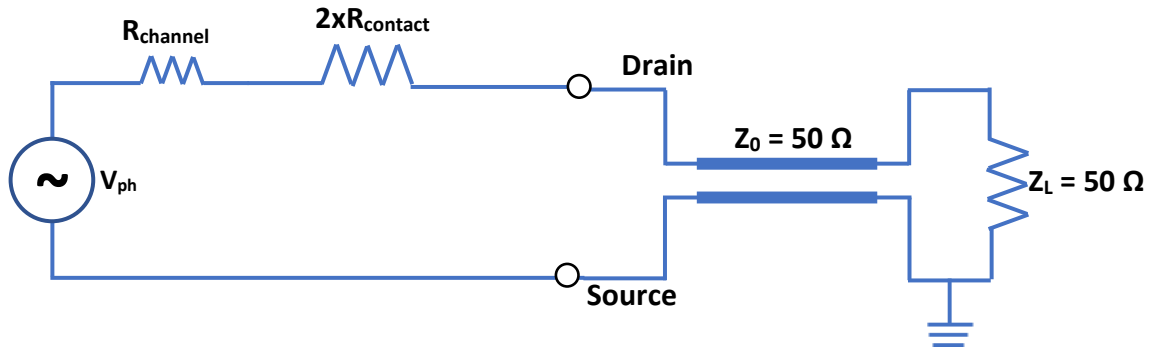

*Fig.S8. Simplified equivalent circuit used in the main text for the series impedance optimization*

**S.8 Series Impedance of the photodetector**

Assuming the simplified model discussed in the previous section, the detector impedance is the geometrical resistance of the graphene channel, i.e. the series resistance $R_{pd}=R_{ch}LW^{-1}+R_cW^{-1}$. Assuming a generated voltage $V_{ph} = R_V P_{av}$, where $R_V$ is the voltage responsivity and $P_{av}$ is the average power of a sinusoidal optical signal, we can write the electrical power $P_L$ transferred to the load $Z_L$ (50 Ω) as[12]:

$$P_L = \frac{1}{2}V_{ph}^2 \frac{Z_L}{(R_{pd}+Z_L)^2} = \frac{1}{2}R_V^2 P_{av}^2 \frac{Z_L}{\left(\frac{2R_c}{W}+\frac{R_{ch}L}{W}+Z_L\right)^2} \tag{s13}$$

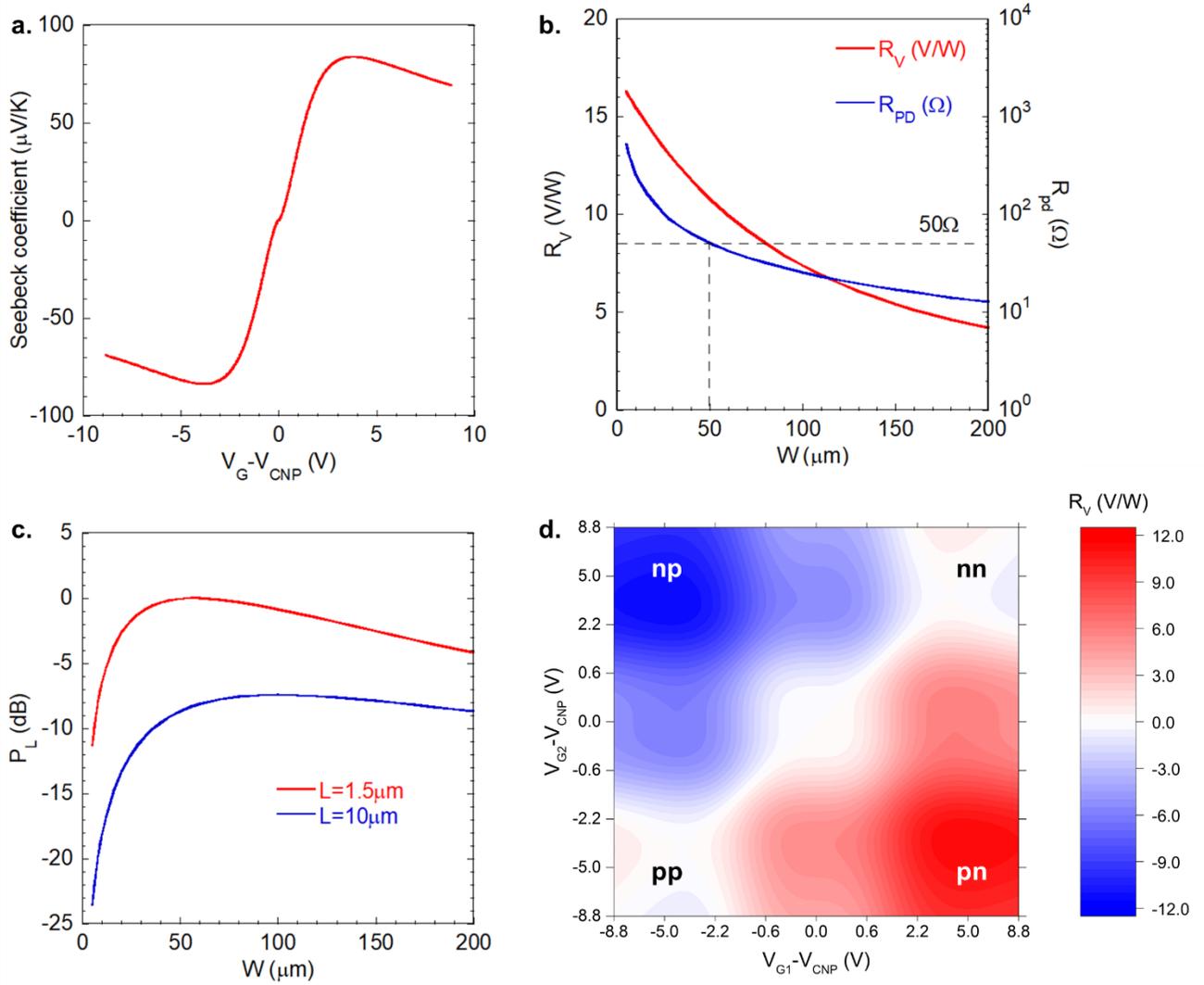

Fig. S9. a. Seebeck coefficient as a function of the gate voltage $V_G$ normalized to charge neutrality voltage $V_{CNP}$. b. Voltage responsivity $R_V$ (red curve) and device resistance $R_{pd}$ (blue curve) versus the channel width W. The horizontal dashed line highlights the 50 Ω resistance value. c. Normalized electrical power $P_L$ transferred from the detector to the input impedance

*of the read-out electronics ($Z_L=50\ \Omega$) as a function of the channel width W. We report the case of a channel length L = 1.5 μm (red curve) and channel length L = 10 μm (blue curve). The two curves are normalized to the maximum power transferred in the case of L = 1.5 μm. d. Six-fold voltage responsivity RV map of the designed PTE photodetector with L = 1.5 μm and W= 50 μm.*

From Equation s13 we show that the optimization of the power transfer is obtained through the optimization of the device length and width. We extracted from simulations the responsivity $R_v$ (for gate voltages corresponding to maximum difference of Seebeck coefficient in the two sides of the junction) as a function of the channel dimensions L and W. This is shown in eq. s13. In Fig. S9.b, we show the evaluated maximum responsivity versus channel width W in the range 5 μm – 200 μm and the corresponding device resistance $R_{pd}$. The photovoltage decreases with the device width W, reaching half of its maximum at about W = 100 μm (Fig. S9.b). The device resistance $R_{pd}$ decreases at increasing channel widths and is 50 Ω for W = 50 μm. The optimal condition for electrical power transfer from the detector to the load is obtained for a device width in the range 50-70 µm (Fig. S9.c). For shorter W, $P_L$ drops because of the larger series resistance of the photodetector. For long contact devices the power transfer is limited by the photovoltage drop. The result is an optimum geometry with L = 1.5 μm and W = 50 μm. In Fig. S9.c we compare $P_L$ for devices with channel length L = 1.5 μm and L = 10 μm. In the latter, non-optimized case, the device resistance is larger, causing a worse impedance matching which results in a lower maximum power transfer ($P_L^{MAX} \approx$ -7 dB). For the optimum design (L = 1.5 μm), the expected voltage responsivity map, as a function of the two gate voltages $V_{G1}$ and $V_{G2}$, is shown in Fig S9.d. The simulation predicts a maximum responsivity of ~10 V W$^{-1}$.